\documentclass[12pt]{iopart}
\usepackage{iopams}
  \usepackage{amssymb}
\catcode`@=11
\newcommand{\overset}[2]{\binrel@{#2}%
  \binrel@@{\mathop{\kern\z@#2}\limits^{#1}}}
\newcommand{\underset}[2]{\binrel@{#2}%
  \binrel@@{\mathop{\kern\z@#2}\limits_{#1}}}
\catcode`@=12

\newfont{\myeu}{eurm10 at 12 pt}
\newfont{\mycal}{eufb10 at 12pt}
\newfont{\bfrak}{eufb10 at 12 pt}
\newcommand{\be}{\begin{equation}}
\newcommand{\ee}{\end{equation}}
\newcommand{\ba}{\begin{eqnarray}}
\newcommand{\ea}{\end{eqnarray}}

\def\bZ{{\bf Z}}
\def\bY{{\bf Y}}
\def\bX{{\bf X}}

\def\bGamma{{\boldsymbol{\Gamma}}}
\def\bH{\mbox{\boldmath$\mathsf{H}$}}

\def\bpsi{{\boldsymbol{\psi}}}

\def\bone{{\mathbf{1}}}
\def\bTheta{{\boldsymbol{\Theta}}}
\def\bP{\mbox{\boldmath$\mathsf{P}$}}

\def\bsigma{{\boldsymbol{\sigma}}}
\def\btau{{\boldsymbol{\tau}}}

\def\bY{{\bf Y}}

\def\vp{{\vphantom{i}}}

\begin{document}
\title[ ]%
{Parafermions in the $\tau_2$ model II}

\author{Helen Au-Yang and Jacques H H Perk}

\address{Department of Physics, Oklahoma State University, \\
145 Physical Sciences, Stillwater, OK 74078-3072, USA}
\ead{perk@okstate.edu, helenperk@yahoo.com}
\begin{abstract}
Many years ago Baxter introduced an inhomogeneous two-dimensional
classical spin model, now called the $\btau_2(t)$ model
with free boundary conditions, and he specialized the resulting
quantum spin-chain Hamiltonian in a special limit to a simple clock
Hamiltonian. Recently, Fendley showed that this clock Hamiltonian
can be expressed in terms of free ``parafermions.'' Baxter followed
this up by showing that this construction generalizes to the more
general $\btau_2(t)$ model, provided some conjectures hold.
In this paper, we will compare the different notations and approaches
enabling us to express the Hamiltonians in terms of projection
operators as introduced by Fendley. By examining the properties of
the raising operators, we are then able to prove the last unproven
conjecture in Baxter's paper left in our previous paper. Thus the
eigenvectors can all be written in terms of these raising operators.
\end{abstract}

\section{Introduction}

In his study of parafermionic spin chains \cite{Fen912,Fen13}, Fendley
was led to consider the simple open spin-chain Hamiltonian introduced
by Baxter \cite{Baxterham,Bax89}, 
\be
{\mathcal{H}}=-\sum_{j=1}^{L}\alpha_j\bX_j-
\sum_{j=1}^{L-1}\gamma_j\bZ^\vp_j\bZ^{-1}_{j+1},
\label{Hamclock}\ee 
which is also equation (B1.5) in \cite{BaxterPf}.\footnote{Equations in
\cite{BaxterPf} are denoted here by prefacing B to the equation number;
those from \cite{Fen13} by adding F in front of their number, and we
preface AP to the equation numbers in \cite{APpf}.} Here $\bX_j$ and
$\bZ_j$ on site $j$ of the chain are copies of the $N$-by-$N$ matrices,
\be\fl
[\bX]_{\sigma,\sigma'}=\delta(\sigma,\sigma'+1),\quad
[\bZ]_{\sigma,\sigma'}=\omega^{\sigma}\delta(\sigma,\sigma'),\quad
\bZ\bX=\omega\bX\bZ,\quad\bZ^N=\bX^N=\bone,
\label{XZ}\ee
generalizing the Pauli matrices $\bsigma^x$ and $\bsigma^z$ to $N>2$
and called $\tau$ and $\sigma$ in \cite{Fen912,Fen13}. Also, in
(\ref{XZ}) we have $\omega=\mathrm{e}^{2\pi\mathrm{i}/N}$ and
$\sigma,\sigma'=0,\cdots,N-1$ in $\mathbb{Z}_N$.

In \cite{Fen13} Fendley succeeded in constructing operators generating
the free parafermions associated with (\ref{Hamclock}), upon which
Baxter generalized \cite{BaxterPf} this construction to the full
inhomogeneous $\tau_2$ model with free boundary conditions from which
Hamiltonian (\ref{Hamclock}) was derived by him
\cite{Baxterham,Bax89,BaxterFun}.
This generalization is of interest as the $\tau_2$ model is an
intermediate \cite{BS,BBP} between the six-vertex model and the
integrable chiral Potts model \cite{AMPTY,BPA,Perk}.

Comparing (\ref{Hamclock}) with (F34) in \cite{Fen13}, we find not
only an overall sign difference, but also the change of $\bZ\to\bZ^{-1}$
corresponding to a left-right reflection of the spin chain.
This is because we are following the conventions of Baxter in
\cite{BaxterPf}, just as we did in our previous paper \cite{APpf},
which the current paper follows up proving the final conjecture in
\cite{BaxterPf} not resolved in \cite{APpf}.\footnote{To facilitate
comparisons with and between the cited papers, we shall outline the
differences in notations and approaches, while also mentioning
equivalences between equations. For more historical context and
citations on parafermions we refer to the introduction of \cite{APpf}.}

\subsection{The $\tau_2$ model with open boundaries
and corresponding Hamiltonian} 
The $\btau_2(t)$ model with cyclic boundary conditions is defined by
Baxter in (B2.6) \cite{BaxterPf} through the transfer matrix between
two rows with spins $\sigma^\vp_0,\cdots,\sigma^\vp_L$ and
$\sigma'_0,\cdots,\sigma'_L$,
\be
\btau_2(t)_{\sigma,\sigma'}=\prod_{j=0}^L
W_j(\sigma^\vp_j,\sigma^\vp_{j+1},\sigma'_{j+1},\sigma'_j),\quad
(\sigma^\vp_{L+1}\equiv\sigma^\vp_0,\quad\sigma'_{L+1}\equiv\sigma'_0),
\label{tau2}\ee
where the nonzero Interaction-Round-a-Face (IRF) weights are given
in (B2.3)\footnote{For the proofs in \cite{APpf} we found it more
convenient to use the equivalent vertex model formulation, see
figure 5 in \cite{BaxterFun}. The IRF formulation used in (\ref{tau2})
and (\ref{weight}) here corresponds to figure 4 in \cite{BaxterFun}.} as
\ba
W_j(\sigma_j,\sigma_{j+1},\sigma_{j+1},\sigma_j)=
b_{2j-1}b_{2j}-\omega^{\sigma_j-\sigma_{j+1}+1}tc_{2j-1}c_{2j},\cr
W_j(\sigma_j,\sigma_{j+1},\sigma_{j+1},\sigma_j-1)=
-\omega td_{2j-1}b_{2j}+\omega^{\sigma_j-\sigma_{j+1}+1}ta_{2j-1}c_{2j},\cr
W_j(\sigma_j,\sigma_{j+1},\sigma_{j+1}-1,\sigma_j)=
b_{2j-1}d_{2j}-\omega^{\sigma_j-\sigma_{j+1}+1}c_{2j-1}a_{2j},\cr
W_j(\sigma_j,\sigma_{j+1},\sigma_{j+1}-1,\sigma_j-1)=
-\omega td_{2j-1}d_{2j}+\omega^{\sigma_j-\sigma_{j+1}+1}a_{2j-1}a_{2j}.
\label{weight}\ea
The transfer matrices $\btau_2(t)$ form a commuting family parametrized
by $t$, irrespective of the choice of the inhomogeneous constants
$a_j,b_j,c_j,d_j$, which are periodic modulo $2L+2$ in $j$, but do not
have to satisfy the chiral Potts curve relations (9) in \cite{BPA}.

In \cite[eq.~73]{BaxterFun} and in (B3.1) Baxter chose $a_{-1}=d_{-1}=0$.
Then, as seen from (\ref{weight}),
$W_0(\sigma^\vp_0,\sigma^\vp_1,\sigma'_1,\sigma'_0)=0$
if $\sigma^\vp_0\ne\sigma'_0$. Therefore, because of periodic
boundary conditions, we must have
$\sigma^\vp_0\equiv\sigma^\vp_{L+1}\equiv\sigma'_0\equiv\sigma'_{L+1}$.
Also, using the functional equations \cite[eq.~47]{BaxterFun}
or (B2.12), Baxter derived
\be
\btau_{2}(t)\btau_{2}(\omega t)\cdots\btau_{2}(\omega^{N-1}t)=
f(t^N)\bone,
\label{taufun}\ee
with $f(x)$ some polynomial and $\bone$ the unit matrix of dimension
$N^{L+1}$. This last statement follows as $z(t)$ in (B2.13)
now vanishes, $\btau_{2}(t)$ is a polynomial in $t$ with matrix
coefficients and (\ref{taufun}) is invariant under $t\to\omega t$.

Baxter specialized further to $a_{-1}=d_{-1}=c_{-1}=c_{2L}=0$,
and $b_{-1}=b_{2L}=1$, see (B3.1) and (B3.4).\footnote{Here we did
not set $b_j=1$ for $0\le j\le2L-1$. As the weights $W_L$ with
$a_{2L}$ and $d_{2L}$ now do not show up in $\btau_2(t)$, we can
set $a_{2L}=d_{2L}=0$ also. The resulting $\btau_2(t)$ is
homogeneous in all its $a,b,c,d$.} Then the relevant boundary
weights become
\ba
W_0(\sigma_0,\sigma_{1},\sigma_{1},\sigma_0)=b_{0},\quad
&W_0(\sigma_0,\sigma_{1},\sigma_{1},\sigma_0-1)=0,\cr
W_0(\sigma_0,\sigma_{1},\sigma_{1}-1,\sigma_0)=d_{0},\quad
&W_0(\sigma_0,\sigma_{1},\sigma_{1}-1,\sigma_0-1)=0,\cr
W_L(\sigma_L,\sigma_{0},\sigma_{0},\sigma_L)=b_{2L-1},\quad
&W_L(\sigma_L,\sigma_{0},\sigma_{0},\sigma_L-1)=-\omega t d_{2L-1}.
\label{weight0}\ea
The two weights $W_L$ in (\ref{weight}) with $\sigma^\vp_0\ne\sigma'_0$
play no role as they are always paired with a vanishing
$W_0\equiv W_{L+1}$ weight. Therefore, from (\ref{weight0}) one
concludes that $\btau_2(t)$ does not depend on $\sigma^\vp_0$ and
$\sigma'_0$. Choosing $\sigma^\vp_0=\sigma'_0=0$ we reduce $\btau_2(t)$
to become a $N^L$-by-$N^L$ matrix. This is how Baxter in \cite{BaxterPf}
made it to be the transfer matrix of a model with free boundary spins
at $j=1$ and $j=L$.\footnote{In Appendix B.2 of \cite{APpf}, we showed
that the object constructed by Fendley in ({F50}) is identical to
this $\btau_2$ in the clock model limit.} From (\ref{weight}) and
(\ref{weight0}) with $W_0$ independent of $t$, it follows that
$\btau_2(t)$ is a polynomial of degree $L$ in $t$,
\be
\btau_2(t)=\sum_{m=0}^L(\omega t)^m\btau_{2,m},\quad
\btau_{2,0}=\btau_2(0)=A_0\bone,\quad
A_0\equiv\prod_{\ell=0}^{2L-1}b_{\ell}.
\label{tauseries}\ee
Therefore, assuming all $b_{\ell}\ne0$, the following
expansion\footnote{Here
${\mathcal{H}}^{(j)}=-{\bf H}^{(j)}$ of (F48), not to be confused
with what is defined in (F81) and (F82).} in powers of $t$,
\be
t\frac d{dt}{\rm ln}\btau_2(t)=
\sum_{m=1}^\infty(\omega t)^m \,{\mathcal{H}}^{(m)},
\quad\btau_2(t)=A_0\exp\Bigg(\sum_{m=1}^\infty
\frac{(\omega t)^m}{m}\,{\mathcal{H}}^{(m)}\Bigg),
\label{tauH}\ee
exists for the inhomogeneous $\btau_2(t)$ model,
with the leading term giving the Hamiltonian
${\mathcal{H}}={\mathcal{H}}^{(1)}$ and all ${\mathcal{H}}^{(m)}$
constituting an infinite set of commuting Hamiltonians. Baxter gave
the explicit form of $\mathcal{H}=A_0^{-1}\btau_{2,1}$ in (B3.22)
using the normalization $b_j\equiv1$.\footnote{From (\ref{weight})
and (\ref{weight0}) one sees that the $\btau_{2,1}$ in
(\ref{tauseries}) allows at most one single block of sites with
$\sigma'_\ell=\sigma^\vp_\ell-1$, as this block can only end
with a weight $W_k$ linear in $t$, forcing all other $W_\ell$
to be constant. In analogy with \cite{Onsager,Kaufman}
each term in $\btau_{2,1}$ has two factors,
one with horizontal interaction proportional to some
$\bZ^\vp_j\bZ_k^{-1}$, followed by the vertical spin flip
$\prod\bX_\ell$ for the entire block. More explicitly, to get
$\btau_{2,1}$, one replaces in $\btau_{2,0}$ on the left either one
$b^\vp_{2j}$ by $d^\vp_{2j}\bX^\vp_{j+1}$ or one $b^\vp_{2j-1}$
by $-c^\vp_{2j-1}\bZ^\vp_j$, and on the right either one
$b^\vp_{2k-1}$ by $-d^\vp_{2k-1}$ or one $b^\vp_{2k}$
by $c^\vp_{2k}\bZ_{k+1}^{-1}$. All intermediate $b^\vp_{2\ell-1}$
are replaced by $a^\vp_{2\ell-1}\bZ^\vp_\ell$ and the $b^\vp_{2\ell}$
by $\omega a^\vp_{2\ell}\bZ_{\ell+1}^{-1}\bX^\vp_{\ell+1}$. Collecting
all $\bX^\vp_\ell$ in the second factor, this is how one can
recover (B3.22).}

From (\ref{taufun}) and (\ref{tauseries}), we conclude that
\be
\btau_{2}(t)\btau_{2}(\omega t)\cdots\btau_{2}(\omega^{N-1}t)=
A_0^N\bone\prod_{j=1}^L(1-r^N_j t^N),
\label{tauN}\ee
where the $L$ parameters $r_j$ are the roots of a degree $NL$
polynomial (B3.16),
\be
s_0 r^{NL}_{j}+s_1  r^{N(L-1)}_{j}+s_2  r^{N(L-2)}_{j}+\cdots+s_L =0,
\quad \hbox{for $j=1,\cdots,L$.}
\label{B3.16}\ee
Thus Baxter obtained all the eigenvalues of the $\btau_2(t)$ matrix,
namely
\be
\tau_{2}(t)=A_0\prod_{j=1}^L(1-r_j \omega^{1+p_j}t),\quad 0\le p_j\le
N-1,\quad 1\le j\le L.
\label{eigentau}\ee
Consequently we have from (\ref{tauH}) also all the $NL$ eigenvalues
of the higher Hamiltonians,
\be
-{\mathcal{H}}^{(m)}|p_1,\cdots,p_L\rangle=
\sum_{j=1}^L (r_j \omega^{p_j})^m|p_1,\cdots,p_L\rangle.
\label{eigenHm}\ee
with $|p_1,\cdots,p_L\rangle$ denoting the corresponding eigenvector.
In section 2 we will discuss what Fendley \cite{Fen13} did to express
such matrices in terms of projection operators.

\subsection{Generalized Jordan--Wigner transform} 

Because of the difference of conventions between \cite{Fen13} and
\cite{BaxterPf}, we have to modify the basic parafermion operators
$\psi_\ell$ defined in (F37). First let us define, in addition to the
$\bX_j$ and $\bZ_j$ defined through (\ref{XZ}), the operators $\bY_j$
as copies of $\bY$ on sites $j$, where
\be\fl
\bY\equiv\omega^{(N-1)/2}\bX^{-1}\bZ=\omega^{(N+1)/2}\bZ\bX^{-1},\quad
\bY^{-1}=\omega^{(1-N)/2}\bZ^{-1}\bX,\quad\bY^N=\bone.
\label{XYZ}\ee
The scalar factors arise, as in the evaluation of $\bY^N$ we have
to commute $\bX^{-1}$ and $\bZ$ exactly $\frac12N(N-1)$ times.
We define the basic parafermions as
\be\fl
{\bpsi}_{2j-2}=
\Bigg(\prod_{\ell=1}^{j-1}\bX^\vp_\ell\Bigg)\bZ^{-1}_{j},\quad
{\bpsi}_{2j-1}=
\Bigg(\prod_{\ell=1}^{j-1}\bX^\vp_\ell\Bigg)\bY^{-1}_{j},\quad
{\bpsi}_{0}={\bGamma}_{0}=\bZ^{-1}_{1},
\label{parafermions}\ee
for $1\leqslant j\leqslant L$. From the commutation relations
of $\bX$, $\bY$ and $\bZ$, it follows that 
\be
{\bpsi}_{j}{\bpsi}_{k}=\omega^{-1}{\bpsi}_{k}{\bpsi}_{j}\quad
\hbox{for $j<k$},\qquad \bpsi_j^N=\bone.
\label{commpf}\ee
If $N=2$, $\omega=-1$, $\omega^{(N-1)/2}=\mathrm{i}$,
and $\bX$, $\bY$ and $\bZ$ become the Pauli matrices $\bsigma^x$,
$\bsigma^y$ and $\bsigma^z$, which are equal to their own inverses.
Then the $\bpsi_\ell$ become the $\bGamma_\ell$ of
Kaufman \cite{Kaufman}.

Hamiltonian (\ref{Hamclock}) is a special case of (B3.23) in
\cite{BaxterPf}, which we rewrite using (\ref{XYZ}) as
\ba\fl
{\mathcal{H}}=&-\sum_{j=1}^{L}\sum_{k=j}^L\omega^{k-j+(N-1)/2}\,
\frac{d_{2j-2}}{b_{2j-2}}
\Bigg(\prod_{\ell=2j-1}^{2k-2}\frac{a_\ell}{b_\ell}\Bigg)
\frac{d_{2k-1}}{b_{2k-1}}\,
\bZ^\vp_j\Bigg(\prod_{\ell=j}^{k-1}\bX^\vp_\ell\Bigg)\bY^{-1}_k
\nonumber\\ \fl
&+\sum_{j=1}^{L-1}\sum_{k=j+1}^L\omega^{k-j-1}\,
\frac{c_{2j-1}}{b_{2j-1}}
\Bigg(\prod_{\ell=2j}^{2k-2}\frac{a_\ell}{b_\ell}\Bigg)
\frac{d_{2k-1}}{b_{2k-1}}\,
\bY^\vp_j\Bigg(\prod_{\ell=j}^{k-1}\bX^\vp_\ell\Bigg)\bY^{-1}_k
\nonumber\\ \fl
&-\sum_{j=1}^{L-1}\sum_{k=j}^{L-1}\omega^{k-j-(N+1)/2}\,
\frac{c_{2j-1}}{b_{2j-1}}
\Bigg(\prod_{\ell=2j}^{2k-1}\frac{a_\ell}{b_\ell}\Bigg)
\frac{c_{2k}}{b_{2k}}\,
\bY^\vp_j\Bigg(\prod_{\ell=j}^{k}\bX^\vp_\ell\Bigg)\bZ^{-1}_{k+1}
\nonumber\\ \fl
&+\sum_{j=1}^{L-1}\sum_{k=j}^{L-1}\omega^{k-j}\,
\frac{d_{2j-2}}{b_{2j-2}}
\Bigg(\prod_{\ell=2j-1}^{2k-1}\frac{a_\ell}{b_\ell}\Bigg)
\frac{c_{2k}}{b_{2k}}\,
\bZ^\vp_j\Bigg(\prod_{\ell=j}^{k}\bX^\vp_\ell\Bigg)\bZ^{-1}_{k+1}.
\label{HBax}\ea
For the special case $N=2$, after rotating $\bZ_\ell\to\bsigma^x_\ell$,
$\bX_\ell\to-\bsigma^z_\ell$ and $\bY_\ell\to\bsigma^y_\ell$, we
recognize a generalized XY-model, like the spin-chain Hamiltonian
that Suzuki introduced \cite{Suzuki,Suzuki2} to commute with the
transfer matrix of the dimer model.

Hamiltonian (\ref{HBax}) may be expressed in terms of the
parafermions (\ref{parafermions}) as
\ba\fl
{\mathcal{H}}=&-\sum_{j=1}^{L}\sum_{m=j}^L\omega^{m-j+(N-1)/2}
\Bigg(\prod_{\ell=2j-1}^{2m-2}\frac{a_\ell}{b_\ell}\Bigg)
\frac{d_{2j-2}d_{2m-1}}{b_{2j-2}b_{2m-1}}
{\bpsi}^{-1}_{2j-2}{\bpsi}_{2m-1}
\nonumber\\ \fl
&-\sum_{j=1}^{L-1}\sum_{m=j}^{L-1}\omega^{m-j}\Bigg[\omega^{-(N+1)/2}
\Bigg(\prod_{\ell=2j}^{2m-1}\frac{a_\ell}{b_\ell}\Bigg)
\frac{c_{2j-1}c_{2m}}{b_{2j-1}b_{2m}}
{\bpsi}^{-1}_{2j-1}{\bpsi}_{2m}
\nonumber\\ \fl
&-\Bigg(\prod_{\ell=2j-1}^{2m-1}\frac{a_\ell}{b_\ell}\Bigg)
\frac{d_{2j-2}c_{2m}}{b_{2j-2}b_{2m}}
{\bpsi}^{-1}_{2j-2}{\bpsi}_{2m}
-\Bigg(\prod_{\ell=2j}^{2m}\frac{a_\ell}{b_\ell}\Bigg)
\frac{c_{2j-1}d_{2m+1}}{b_{2j-1}b_{2m+1}}
{\bpsi}^{-1}_{2j-1}{\bpsi}_{2m+1}\Bigg],
\label{Hamtau}\ea
setting $k=m+1$ in the second term of (\ref{HBax}) and $k=m$ in
the other three. Thus we find that $\mathcal{H}$ is quadratic in
the parafermions, just like operators in the Onsager algebra
\cite{Onsager} and the generalized XY-model are quadratic in
fermions \cite{JhaV}.
If, as Baxter did in (B3.25), we set $a_j=0$ for $j=1,\cdots,2L-2$
in (\ref{Hamtau}), only the terms with $m=j$ in the first two lines
of (\ref{Hamtau}) survive, the empty products over $\ell$ being
equal to 1. Then (\ref{Hamtau}) reduces to the Hamiltonian below
(F37) in \cite{Fen13} corresponding to (F34), which is also Baxter's
special clock Hamiltonian (\ref{Hamclock}).

\subsection{Raising Operators} 
Inspired by Fendley's paper \cite{Fen13}, Baxter defined in (B4.2)                                                                                                                                                                                                                                                                                                                                                                                                       
\be
\bGamma_0=\bZ_1^{-1},\quad  
\bGamma_{j+1}=(\omega^{-1}-1)^{-1}
({\mathcal{H}}\bGamma_j-\bGamma_j{\mathcal{H}}),\quad (j\geqslant0),
\label{gamma0j}\ee
which is almost the same as (F80). Using $\bGamma_0=\bpsi_0$,
(\ref{commpf}) and (\ref{Hamtau}), it is straightforward to show that
\be\fl
\bGamma_1=\frac{d_0}{b_0}\left[\sum_{m=1}^L\omega^{m+(N-1)/2}
\Bigg(\prod_{\ell=1}^{2m-2}\frac{a_\ell}{b_\ell}\Bigg)
\frac{d_{2m-1}}{b_{2m-1}}\bpsi_{2m-1}-\sum_{m=1}^{L-1}\omega^m
\Bigg(\prod_{\ell=1}^{2m-1}\frac{a_\ell}{b_\ell}\Bigg)
\frac{c_{2m}}{b_{2m}}\bpsi_{2m}\right],
\label{gamma1}\ee
which is rather complicated. Nevertheless, using (\ref{commpf})
again, we can easily show
\be
\bGamma_0\bGamma_1=\omega^{-1}\bGamma_1\bGamma_0.
\label{commg01}\ee

Based on numerical evidence, Baxter found that the infinite
sequence of the $\bGamma_j$ truncates, as he conjectured that
the $\bGamma$ matrices satisfy the equation
\be
s_0 \bGamma_{NL+j}+s_1 \bGamma_{N(L-1)+j}+\cdots+s_L
\bGamma_{j}=0,\quad \hbox{for $j=0$,}
\label{B4.3}\ee
with the same $s_\ell$ as in (\ref{B3.16}), which is also (B3.16).
In \cite{APpf}, we have proven that this equation holds for any
nonnegative $j$. Fendley on the other hand introduced a different
basis in (F84) by making certain subtractions so that the
iteration terminates as shown in (F88). Since the Hamiltonian
(\ref{Hamtau}) is more complicated than (\ref{Hamclock}),
his method may not work in the general case. 

We can use (\ref{gamma0j}) to express the commutators
$[\mathcal{H},\bGamma_j]$ in terms of $\bGamma_{j+1}$
for $0\leqslant j<NL$ and use (\ref{B4.3}) to eliminate
$\bGamma_{NL}$. Thus we recover (B4.11),
\be
{\mathcal{H}}\bGamma_j-\bGamma_j{\mathcal{H}}=
(\omega^{-1}-1)\sum_{k=0}^{NL-1}h_{jk}\bGamma_k,
\label{B4.11}\ee
where
\ba 
h_{ij} =\delta_{i+1,j},\quad
(0\leqslant i\leqslant NL-2,\;0\leqslant j\leqslant NL-1);
\nonumber\\
h_{NL-1,mN}=-s_{L-m}/s_0,\quad(0\leqslant m\leqslant L-1),
\nonumber\\
h_{NL-1,j}=0,\quad (j\not\equiv 0\hbox{ mod }N).
\label{hij}\ea
In (B4.10), Baxter denoted this $NL\times NL$ matrix with
elements $h_{jk}$ by $\bH$, (not to be confused with the
Hamiltonian $\mathcal{H}$), and he showed that its characteristic
polynomial is
\be
|\bH-\lambda{\bf I}|=s_0\lambda^{NL}+s_1\lambda^{N(L-1)}
+\cdots+s_{L-1}\lambda^{N}+s_L.
\label{spol}\ee
Comparing with (\ref{B3.16}), we find the $NL$ roots to be
$\lambda_i=r_k\omega^p$, with $1\leqslant k\leqslant L$ and
$0\leqslant p\leqslant N-1$.

In section 5 of \cite{APpf}, we calculated the eigenvectors
of $\bH$, which form the columns of the matrix $\bP$
diagonalizing $\bH$,
\be
\bP^{-1}\bH\bP=\bH_d,
\ee
and we found that $\bP$ is the Vandermonde matrix
\be
\bP=\left[\begin{array}{ccccc}
1&1&1&\cdots&1\\
\lambda_1&\lambda_2&\lambda_3&\cdots&\lambda_{NL}\\
\lambda^2_1&\lambda^2_2&\lambda^2_3&\cdots&\lambda^2_{NL}\\
\lambda^3_1&\lambda^3_2&\lambda^3_3&\cdots&\lambda^3_{NL}\\
\vdots&\vdots&\vdots&\cdots&\vdots\\
\lambda^{NL-1}_1&\lambda^{NL-1}_2&\lambda^{NL-1}_3
&\cdots&\lambda^{NL-1}_{NL}
\end{array}\right].
\label{Vande}\ee
In order to be consistent with notations in \cite{BaxterPf} we choose
its matrix indices as follows,
\be
P_{ij}=\lambda_j^{\;i},\quad \mbox{with $0<i<NL-1$, $1<j<NL$}.
\label{Vande2}\ee
According to Prony's 1795 result \cite{Prony,Muir,APsu2}, the elements
$(\bP^{-1})_{jk}$ of its inverse are the coefficients
of the polynomials $f_j(z)$ given by
\be
f_j(z)=\prod_{i=1,i\ne j}^{NL}\frac{z-\lambda_i}{\lambda_j-\lambda_i}
=\sum_{k=0}^{NL-1}(P^{-1})_{jk}z^k, \quad \hbox{satisfying}
\quad f_j(\lambda_i)=\delta_{ji}.
\label{Pinverse}\ee
In (B4.17) Baxter then defined new raising operators,
\be
\widehat\bGamma_i=\sum_{j=0}^{NL-1}P^{-1}_{ij}\bGamma_j
\label{wG}\ee
so that (\ref{B4.11}) becomes (B4.18),
\be
{\mathcal{H}}\widehat\bGamma_j-\widehat\bGamma_j{\mathcal{H}}
=(\omega^{-1}-1)\lambda_j\widehat\bGamma_j.
\label{B4.18}\ee
\section{Projection operators}
We have shown in appendix B of \cite{APpf} that the inverse Vandermonde
in (\ref{Pinverse}) is related to the inverse of Fendley's Vandermonde
matrix ${\mathcal{X}}$ on page 28 of \cite{Fen13}, namely
\be
P^{-1}_{i,\ell N+q}=P^{-1}_{(p,k),\ell N+q}=
\frac 1N({\mathcal{X}}^{-1})_{k,\ell}(r_k\omega^p)^{-q},\quad i=kN+p,
\label{PXi}\ee
which is (AP$\,$B.11) in \cite{APpf}.
Consequently, combining (F100) and (F103) and generalizing the result
to our $\btau_2$ case, we find that the projection operator is 
\be
{\mathcal{P}}_{\omega^p,k}=-\sum_{\ell=0}^{L-1}\sum_{q=0}^{N-1}
P^{-1}_{p,k;\ell N+q}{\mathcal{H}}^{(\ell N+q)}, 
\label{Proj}\ee
where we use $\mathcal{P}$ for the projection operator in order
to distinguish it from the Vandermonde matrix $\bP$. Using 
\be
\bP\cdot\bP^{-1}=1,
\ee
where $\bP$ is given in (\ref{Vande}), and letting
$\lambda_i=r_k\omega^p$ and $i=kN+p$, we find
\be \sum_{i=1}^{NL}\lambda_i^m P^{-1}_{i,\ell N+q}=
\sum_{k=1}^{L}\sum_{p=0}^{N-1}(r_k\omega^p)^m P^{-1}_{(p,k),\ell N+q}
=\delta_{m,\ell N+q}.
\ee
Multiplying (\ref{Proj}) by $\lambda^m_i$ and summing over all $i$,
we find
\be
{\mathcal{H}}^{(m)}=-\sum_{k=1}^{L}\sum_{p=0}^{N-1}
(r_k\omega^p)^m{\mathcal{P}}_{\omega^p,k},
\label{H-P}\ee
which is the same as (F105), but now generalized to the full
$\btau_2(t)$ model with free boundaries.
Since the ${\mathcal{H}}^{(m)}$ commute with one another,
the projection operators in (\ref{Proj}) must also. Thus
\be
[{\mathcal{P}}_{\omega^p,k},{\mathcal{P}}_{\omega^q,\ell}]=0.
\label{comproj}
\ee
In (\ref{eigenHm}) we introduced the basis of $N^L$ eigenstates
$\{|n_1,n_2,\cdots,n_L\rangle, (n_j\in\mathbb{Z}_N)\}$ on which
the ${\mathcal{H}}^{(m)}$ act as
\be
{\mathcal{H}}^{(m)}|n_1,n_2,\cdots,n_L\rangle=
-\sum_{k=1}^{L}(r_k\omega^{n_k})^m\,|n_1,n_2,\cdots,n_L\rangle.
\label{eigenm}\ee
Substituting (\ref{H-P}) into (\ref{eigenm}), we find that
\be
{\mathcal{P}}_{\omega^p,k}|n_1,n_2,\cdots,n_L\rangle=
\delta_{p,n_k}|n_1,n_2,\cdots,n_L\rangle,
\label{proj1}\ee
which shows that
\be
{\mathcal{P}}^2_{\omega^p,k}= {\mathcal{P}}_{\omega^p,k},\quad
{\mathcal{P}}_{\omega^p,k}{\mathcal{P}}_{\omega^q,k}=
\delta_{p,q}{\mathcal{P}}_{\omega^p,k},\quad
\sum_{p=0}^{N-1}{\mathcal{P}}_{\omega^p,k}=\bone.
\label{proj2}\ee
These properties show that the ${\mathcal{P}}_{\omega^p,k}$
are indeed projection operators, agreeing with what Fendley found
in \cite{Fen13} for the special case (\ref{Hamclock}). Next we set
$\lambda_j=r_\ell\omega^q$ and $j=\ell N+q$ in (\ref{B4.18}),
so that $\widehat\bGamma_j\equiv\widehat\bGamma_{q,\ell}$.
Then using (\ref{H-P}) with $m=1$, we have
\be
\sum_{k=1}^{L}\sum_{p=0}^{N-1}(r_k\omega^p)
[{\mathcal{P}}_{\omega^p,k}\widehat\bGamma_{q,\ell}
-\widehat\bGamma_{q,\ell}{\mathcal{P}}_{\omega^p,k}]=
r_\ell(\omega^{q-1}-\omega^q)
\widehat\bGamma_{q,\ell}.
\ee
This implies the relation, 
\be
[{\mathcal{P}}_{\omega^p,k}\widehat\bGamma_{q,\ell}
-\widehat\bGamma_{q,\ell}{\mathcal{P}}_{\omega^p,k}]=
\delta_{k,\ell}(\delta_{p,q-1}-\delta_{p,q})
\widehat\bGamma_{q,\ell},
\label{F106}\ee
in agreement with the equation above (F106) in \cite{Fen13}.
In the derivation of (\ref{F106}) we used that
$\widehat\bGamma_{q,\ell}$ only acts on the $n_\ell$ in the
eigenstate $|n_1,\cdots,n_L\rangle$, as was shown by
Baxter \cite{BaxterPf} using (B4.21) following from
(B4.19),\footnote{Some steps in the derivation in \cite{BaxterPf}
were found from numerical work and proved in \cite{APpf}.}
which we can rewrite as
\be
(1-r_\ell\omega^{q}\,t)\btau_2(t)\widehat\bGamma_{q,\ell}
=(1-r_\ell\omega^{q+1}\,t)\widehat\bGamma_{q,\ell}\btau_2(t)
\label{B4.21}\ee
and from which (\ref{B4.18}) also follows as the first
nontrivial term in the expansion in powers of $\omega t$.
The extra $t$-dependence means that we can forget about
complications due to accidental degeneracies. The ratio of the
coefficients in (\ref{B4.21}) is the ratio of two unique eigenvalues
of $\btau_2(t)$, so that $\widehat\bGamma_{q,\ell}$ has its only
one nonzero matrix element between the two corresponding eigenvectors
\cite[section 4.3]{BaxterPf}, raising $n_\ell=q-1$ to $q$.

In terms of the above projection operators, and using their properties
(\ref{proj1}) and (\ref{proj2}), we find from (\ref{eigentau})
\be\fl
\btau_2(t)=A_0\prod_{k=1}^{L}\prod_{p=0}^{N-1}
(\bone-r_k\omega^{1+p}t\,{\mathcal{P}}_{\omega^p,k})=
A_0\prod_{k=1}^{L}\Bigg(\bone-\omega t\sum_{p=0}^{N-1}
r_k\omega^{p}\,{\mathcal{P}}_{\omega^p,k}\Bigg).
\label{tau-P}\ee
This agrees with (AP75) in \cite{APpf} only if one identifies
\be
\mathbf{u}_k=\sum_{p=0}^{N-1}r_k\omega^{p}\,{\mathcal{P}}_{\omega^p,k}.
\ee

Finally, as shown in \cite{BaxterPf, APpf}, the only non-vanishing
elements of $\widehat\bGamma_{p,k}$ are
\ba
\langle n_1,\cdots,{\overset {k}{p}},\cdots, n_L|\widehat\bGamma_{p,k}
|n_1,\cdots,{\overset {k}{p\!-\!1}},\cdots,n_L\rangle
\nonumber\\
\qquad=\langle n_1,\cdots,{\overset {k}{p}},\cdots, n_L|
\bGamma_0|n_1,\cdots,{\overset {k}{p\!-\!1}}\cdots,n_L\rangle,
\label{Gright2}\ea
see (AP86) and (AP95) for example. Each eigenstate is represented
by $L$ integers $n_1,\cdots,n_k,\cdots n_L$ with $n_k=0,\cdots, N-1$.
The raising operator $\widehat\bGamma_{p,k}$ raises the value
$n_k$ by one (mod $N$) if $n_k=p-1$ leaving the other $L-1$
integers unchanged; if $n_k\ne p-1$, it kills the eigenstate.

\section{Proof of (B5.4) or (AP96)}

However, there is a much easier way to prove (\ref{Gright2})
and to generalize it using the Vandermonde matrix (\ref{Vande}).
Let $\{n^\vp_\ell\}^\vp_k$ be the set $\{n'_\ell\}$ with
$n'_\ell=n^\vp_\ell$ for $\ell\ne k$ and $n'_k=n^\vp_k-1$ (mod $N$).
Then from Baxter's argument \cite{BaxterPf}---see text around
(\ref{B4.21})---we know that
$\langle\{n^\vp_\ell\}|\widehat\bGamma_{i}|\{n'_\ell\}\rangle=0$,
if $\{n'_\ell\}\ne\{n^\vp_\ell\}^\vp_i$. Applying the
Vandermonde matrix $\bP$, we also find
$\langle\{n^\vp_\ell\}|\bGamma_{j}|\{n'_\ell\}\rangle=0$
for all $j$, if $\{n'_\ell\}\ne\{n^\vp_\ell\}^\vp_k$
for all $k$. More precisely,
\be
\langle\{n_\ell\}|\bGamma_{j}|\{n_\ell\}_k\rangle=
(\lambda_{kN+\ell})^j
\langle\{n_\ell\}|\widehat\bGamma_{kN+\ell}|\{n_\ell\}_k\rangle,
\quad \lambda_{kN+\ell}=r_k\omega^{n_\ell}.
\label{gammaj0}\ee
Thus the $\bGamma_{j}$ can only have elements
corresponding to raising one $n_\ell$ by 1, so that the only
nonzero elements are
\be  \langle n_1,\cdots,n_\ell,\cdots,n_L|\bGamma_{j}
|n_1,\cdots,n_\ell-1,\cdots,n_L\rangle\ne0, \quad \ell=1,\cdots,L.
\label{gamma0}\ee
Unlike the raising operator $\widehat\bGamma_{k,p}$ which
only can raise $n_k=p-1$ by one, we find
$\bGamma_{j}$ can raise any $n_\ell$ by one for any $\ell$.

Now we use (\ref{commg01}) to prove (AP96). We write
\ba\fl
0=\langle n_1,\cdots,{\overset {k}{p}},\cdots,
{\overset {\ell}{q}},\cdots,n_L
|(\bGamma_0\bGamma_1-\omega^{-1}\bGamma_1\bGamma_0)
|n_1,\cdots,{\overset {k}{p-1}},\cdots, {\overset {\ell}{q-1}},
\cdots,n_L\rangle
\cr\cr\fl
=\sum_{\{n'_i\}}\bigg[\langle n_1,\cdots,{\overset {k}{p}},\cdots,
{\overset {\ell}{q}},\cdots,n_L|\bGamma_0|\{n'_i\}\rangle
\langle\{n'_i\}|\bGamma_1|n_1,\cdots,{\overset {k}{p-1}},\cdots,
{\overset {\ell}{q-1}},\cdots,n_L\rangle
\nonumber\\ \fl
-\omega^{-1}\langle n_1,\cdots,{\overset {k}{p}},\cdots,
{\overset {\ell}{q}},\cdots,n_L|\bGamma_1|\{n'_i\}\rangle
\langle\{n'_i\}|\bGamma_0|n_1,\cdots,{\overset {k}{p-1}},\cdots,
{\overset {\ell}{q-1}},\cdots,n_L\rangle\bigg].
\label{AP96}\ea
From (\ref{gamma0}), we find only two possibilities for the
summand to be nonvanishing: either $n'_k=p-1$ and $n'_i=n_i$
for $i\ne k$, or $n'_\ell=q-1$ and $n'_i=n_i$ for $i\ne \ell$.
Furthermore, we find from (\ref{gammaj0}) that, 
\ba\fl
\langle n_1,\cdots,{\overset {k}{p}},\cdots,
{\overset {\ell}{q-\epsilon}},\cdots,n_L|\bGamma_1|n_1,\cdots,
{\overset {k}{p-1}},\cdots, {\overset {\ell}{q-\epsilon}}
,\cdots,n_L\rangle
\nonumber\\ \fl\qquad
=r_k\omega^p\langle n_1,\cdots,{\overset {k}{p}},\cdots,
{\overset {\ell}{q-\epsilon}},\cdots,n_L|\bGamma_0|n_1,\cdots,
{\overset {k}{p-1}},\cdots, {\overset {\ell}{q-\epsilon}},
\cdots,n_L\rangle,
\nonumber\\ \fl
\langle n_1,\cdots,{\overset {k}{p-\epsilon}},\cdots,
{\overset {\ell}{q}},\cdots,n_L|\bGamma_1|n_1,\cdots,
{\overset {k}{p-\epsilon}},\cdots, {\overset {\ell}{q-1}},
\cdots,n_L\rangle
\nonumber\\ \fl\qquad
=r_\ell\omega^q\langle n_1,\cdots,
{\overset {k}{p-\epsilon}},\cdots, {\overset {\ell}{q}},
\cdots,n_L|\bGamma_1|n_1,\cdots,{\overset {k}{p-\epsilon}},
\cdots,{\overset {\ell}{q-1}},\cdots,n_L\rangle,
\label{gamma10}\ea
for $\epsilon=0,1$. Substituting (\ref{gamma10}) into
(\ref{AP96}), we prove the identity in (AP96) and therefore also
(B5.4), which generalizes (F111).

Finally, we can construct all the eigenvectors by applying
the raising operators on the `ground state' with
$n_1=n_2=\cdots=n_L=0$ and denoted by
\be
|\{0\}\rangle=|0,0,\cdots,0\rangle.
\label{ground}\ee
Let the ordered product of $p$ raising operators
$\widehat\bGamma_{q,k}$ in descending order of $q$ be
\be
\bTheta_{p,k}=\widehat\bGamma_{p,k}\widehat\bGamma_{p-1,k}
\cdots\widehat\bGamma_{2,k}\widehat\bGamma_{1,k},
\qquad \bTheta_{0,k}={\bf 1}.
\label{Theta}\ee
Any eigenvectors can be obtained as
\be |\{n_i\}\rangle=|n_1,n_2,\cdots,n_L\rangle=C(\{n_i\})
\bTheta_{n_1,1}\bTheta_{n_2,2}\cdots\bTheta_{n_L,L}|\{0\}\rangle.
\label{eigenstate}\ee
Alternatively, we can use
\be\fl
\widehat\bTheta_{p,k}\equiv\Bigg(\sum_{q=0}^{N-1}
\widehat\bGamma_{q,k}\Bigg)^p,\quad
\widehat\bTheta_{N,k}=
\sum_{q=0}^{N-1}\widehat\bGamma_{N+q,k}\widehat\bGamma_{N-1+q,k}
\cdots\widehat\bGamma_{2+q,k}\widehat\bGamma_{1+q,k},
\ee
identifying $\widehat\bGamma_{N+q,k}\equiv\widehat\bGamma_{q,k}$.
If $\widehat\bTheta_{N,k}\propto\bone$, as in proposal (F108), we can
call the $\widehat\bTheta_{1,k}$ cyclic raising (or shift) operators.
One may consult section 6.3 of \cite{Fen13} for further discussion
related to the special Hamiltonian (\ref{Hamclock}).

\section{Summary}
In this paper we presented some new results for the inhomogeneous
$\btau_2$ model with open boundary conditions and its associated
Hamiltonians. In section 1 we have given an introduction including
several formulae from papers of Baxter \cite{BaxterPf}, Fendley
\cite{Fen13} and ourselves \cite{APpf} that are needed to make the
present paper somewhat self-contained. We added new details and
discussions and we discussed the differences in notations and symbols
between the papers stemming in part from differences in conventions
between \cite{Fen13} and \cite{BaxterPf}. We reviewed the
eigenvalue spectrum and quantum numbers, and also
the two sets of operators $\bGamma_j$ and $\widehat\bGamma_j$.

In (\ref{Proj}) of section 2 we introduced the complete set of
projection operators ${\mathcal{P}}_{\omega^p,k}$ defined in terms of
the higher Hamiltonians ${\mathcal{H}}^{(m)}$, in full analogy with
(F100) and (F103) for the special clock model; only we used
${\mathcal{H}}^{(m)}$ instead of the $-{\bf H}^{(m)}$ of Fendley.
As a consequence, in (\ref{H-P}) and (\ref{tau-P}), the Hamiltonians
${\mathcal{H}}^{(m)}$ and the $\btau_2(t)$ matrix are all
expressed in terms of the projection operators.

We then showed in section 3, applying the Vandermonde matrix, that
the elements of the operators
$\langle\{n'_i\}|\bGamma_{j}|\{n_i\}\rangle$ can be non-zero if
and only if any one of $L$ integers, say $n_k$, increase by one,
$n'_k=n_k+1$. Finally, we proved conjecture (B5.4) or equivalently
(AP96), generalizing (F111) and giving us the commutation relation
for $\widehat\bGamma_{p,k}$ and $\widehat\bGamma_{q,\ell}$
with $k\ne\ell$.



\end{document}